\documentclass{article}
\usepackage[utf8]{inputenc}
\usepackage{graphicx,wrapfig}
\usepackage{amsmath}
\usepackage{chngcntr}
\usepackage{multicol}
\usepackage[english]{babel}
\usepackage{multirow}
\usepackage{url}
\usepackage[square,numbers]{natbib}
\usepackage{lscape}
\usepackage{longtable}
\usepackage{amsfonts}
\usepackage{amssymb}
\usepackage{siunitx}
\usepackage{appendix}
\usepackage{listings}
\usepackage{float}
\usepackage{subcaption}
\usepackage{chngcntr}
\counterwithin{figure}{section}
\usepackage{fancyhdr}
\makeindex
\usepackage[intoc]{nomencl}
\makenomenclature

\title{Photothermal and Thermo-optical Effects in 3D Arrays of Dielectric and Plasmonic Nanoantennas}
\author{Alfredo Naef, Ted V. Tsoulos, Giulia Tagliabue}
\date{Laboratory of Nanoscience for Energy Technologies (LNET), Ecole Polytechnique Federale de Lausanne (EPFL)}

\begin{document}
\maketitle
\counterwithout{figure}{section}
\begin{figure}[h!]
    \centering
    \includegraphics[width=8cm]{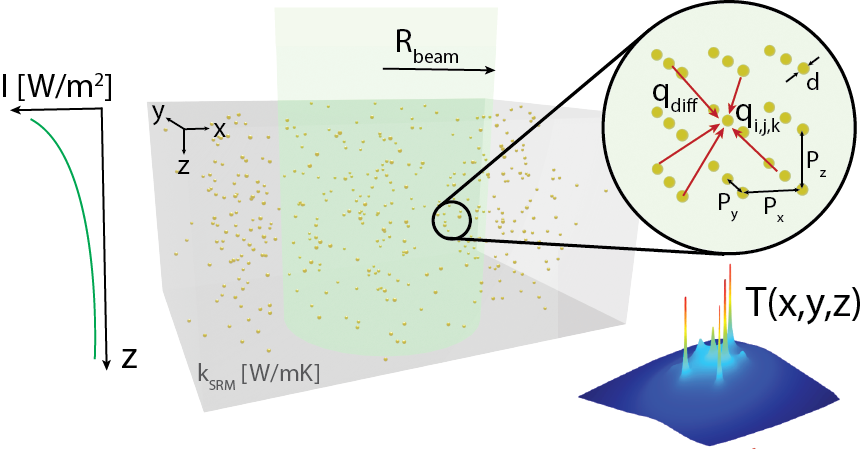}
\end{figure}

\begin{abstract} 
Thermonanophotonics, i.e. the study of photothermal effects in optical nanoantennas, has recently attracted growing interest. While thermoplasmonic structures enable a broad range of applications, from imaging and optofluidics devices to medical and photochemical systems, dielectric nanoantennas open new opportunities for thermo-optical modulation and reconfigurable metasurfaces. However, computing both photo-thermal and thermo-optical effects in large arrays of nanoantennas remains a challenge. In this work, we implement a fast numerical method to compute the temperature increase of multi-dimensional arrays of optical antennas embedded in a uniform medium, accounting for self-heating, collective heating as well as thermo-optical effects. In particular, we demonstrate scalable computation of temperature in 3D networks with $10^5$ particles in less than 1 hour. Interestingly, by explicitly considering the role of discrete nanoparticles on light attenuation and photothermal conversion, this approach enables the optimization of complex temperature profiles in 3D arrays. Importantly, we compute for the first time the impact of thermo-optical effects beyond the single nanoantenna. Our results show that collective heating contributions amplify these effects in multi-dimensional arrays of both Silicon and Gold nanospheres, highlighting the importance of considering them in photothermal calculations. Overall, the proposed method opens new opportunities for the rapid assessment of complex photothermal effects in arrays of optical nanoantennas, supporting the development of advanced thermonanophotonic functionalities.

\textbf{keywords}: thermonanophotonics, thermoplasmonic, nanoantennas, dielectric nanoresonators, thermo-optical effects 

\end{abstract}
\clearpage
\twocolumn

\section{Introduction}
\paragraph{}
The intense and strongly localized photothermal effects generated by resonant illumination of nanoantennas have been attracting a fast-growing interest in the last two decades. Nanoheat sources based on plasmonic nanoparticles have indeed been employed in a growing number of applications such as photothermal cancer therapy \cite{nam_ph-induced_2009}, nanosurgery \cite{boulais_plasmonics_2013}, drug delivery \cite{ahmad_metallic_2010}, photoacoustic imaging \cite{bayer_photoacoustic_2013}, phase transitions \cite{neumann_solar_2013}\cite{richardson_experimental_2009}, optofluidics \cite{liu_optofluidic_2006} and photochemistry \cite{baffou_simple_2020}. Recent studies have also shown that thermoplasmonic heating of bulk materials can have significant advantages for the synthesis of polymer powders \cite{powell_white_2018}, for curing resins \cite{roberts_plasmonic_2019}, for the self-healing of polymers and glasses \cite{zhang_polymers_2013}\cite{zhang_light-healable_2014}\cite{fan_plasmonic_2019}  as well as for contactless actuation \cite{ding_light-induced_2016} \cite{grzelczak_stimuli-responsive_2019} \cite{parreira_remotely_2020}. Interestingly, photothermal effects in dielectric nanoantennas (e.g. silicon) have also become the subject of intense research recently, thanks to the possibility of achieving photothermal modulation of the optical response via thermo-optical effects \cite{tsoulos_self-induced_2020}\cite{archetti_thermally_2022}\cite{ryabov_nonlinear_2022}\cite{iyer_uniform_2018}\cite{butakov_switchable_2018}. Overall, thermonanophotonics is thus a growing research area that offers promising opportunities for a wide range of applications \cite{lewi_thermally_2019}\cite{zograf_all-dielectric_2021}. 

\begin{figure}[h!]
    \centering
    \includegraphics[width=8cm]{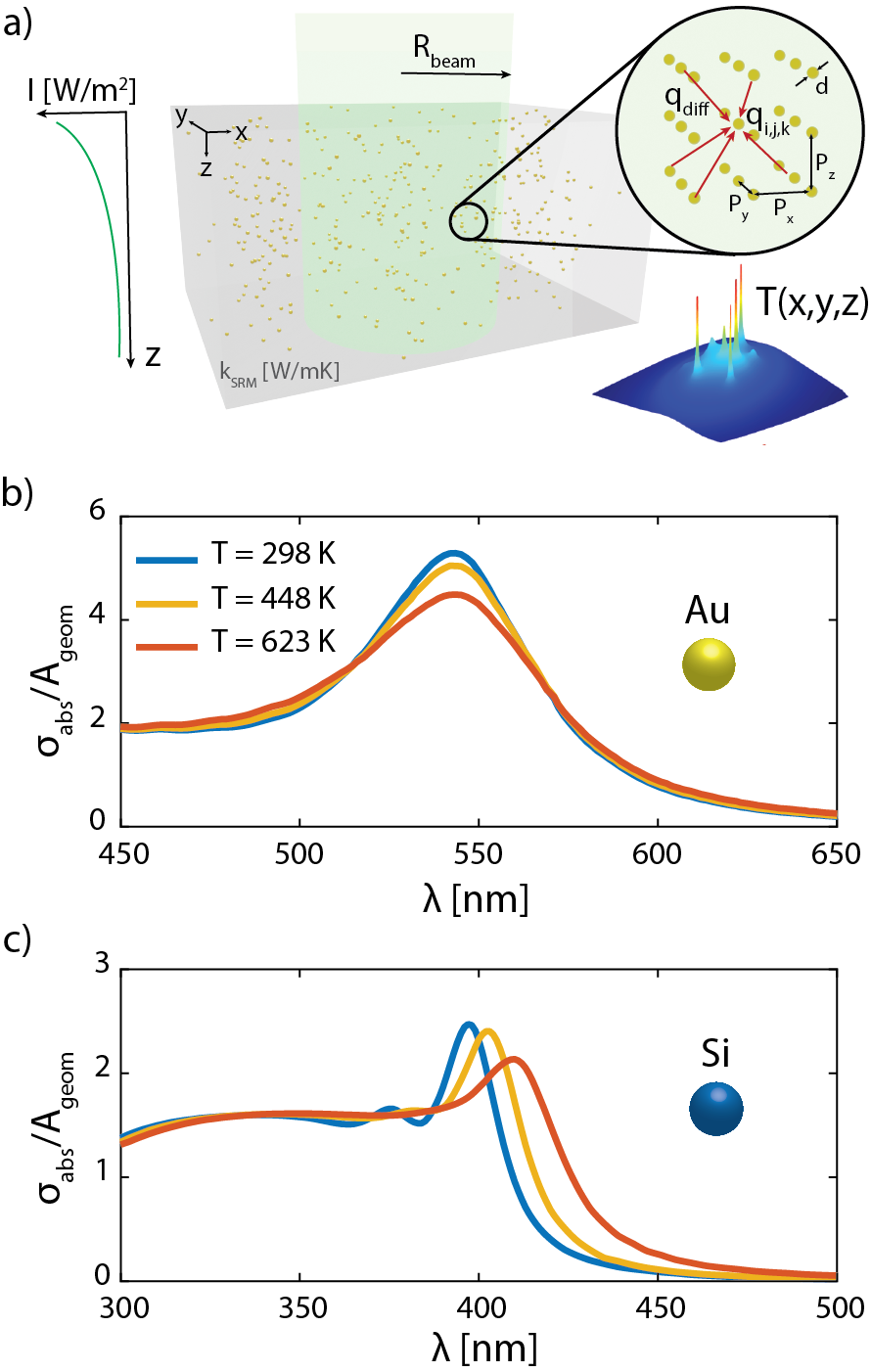}
    \caption{System Definition and Input Optical Properties \textbf{a)} A homogeneous 3D dispersion of optical nanoantennas (diameter $d$) embedded in a non-absorbing solid medium (thermal conductivity $k_{SRM}$) is approximated with a cartesian network of nanoantennas (pitch $P_x, P_y, P_z$). The network is excited by a light beam of radius $R_{beam}$ and uniform intensity $I_0$, propagating along the z-direction. Accounting for the intensity decay ($I(z)$) and thermo-optical effects, the model determines both self-heating ($q_{i,j,k}$) and collective heating ($q_{diff}$) contributions to the temperature of each nanoabsorber, finally interpolating the spatial temperature profile. \textbf{b)} Thermo-optical effect on the absorption efficiency of a gold nanosphere with $R=26nm$ and \textbf{c})  a silicon nanosphere with $R=33nm$ .}
    \label{fig:abstract}
\end{figure}

However, computing and engineering a specific temperature profile within a multi-dimensional, and in particular three dimensional (3D), array of resonant nanoparticles remains challenging. Indeed, even in the absence of strong light scattering or electromagnetic coupling between neighboring nanoantennas, the final temperature depends on the interplay of optical absorption in each nanoparticle, i.e. self-heating, and heat diffusion, i.e. collective heating. In 3D systems, self-heating is also affected by light attenuation (\textbf{Figure \ref{fig:abstract}a}). This, in turn, depends on the optical properties of the nanoantennas and their spatial arrangement. Hence, the photothermal response of the entire array, which can consist of thousands of elements, must be in principle computed. Numerical methods (e.g. COMSOL) face significant challenges in this regard and cannot be used for fast screening or optimization. In fact, micro- or millimeter-scale computational volumes containing nanometer scale heat sources result in large meshes and long computational times. A Green-dyadic method \cite{baffou_thermoplasmonics_2010} was proposed and widely applied, in particular for calculating the temperature profile of two-dimensional (2D) arrays of plasmonic nanoparticles. Also, analytical expressions were derived to estimate the global peak temperature increase within multi-dimensional arrays \cite{richardson_experimental_2009}\cite{baffou_photoinduced_2013} while a non-dimensional parameter was defined to rapidly identify photothermal regimes dominated by self-heating or collective heating effects \cite{baffou_photoinduced_2013}\cite{baffou_thermoplasmonics_2017}, resulting in highly peaked and smooth temperature profiles, respectively. However, despite interesting applications   \cite{roberts_plasmonic_2019}\cite{hogan_nanoparticles_2014}\cite{richardson_experimental_2009}, 3D arrays have only been optimized for cases dominated by collective heating effects. In fact, in this condition the Beer-Lambert law can be used to estimate light attenuation while a volumetric heat source \cite{waiun_parametric_2020} resolves the complexity of accounting for discrete light absorbers. Yet, this approach cannot be employed when self-heating effects are significant (e.g. strong light intensity, sparse nanoparticles). Thus, a broader analysis of the photothermal behaviour of 3D arrays is still missing. Importantly, thermo-optical effects, which depend on the local temperature, alter the absorption cross-section of each nanoantenna in an array differently \textbf{(Figure \ref{fig:abstract}b,c)}, affecting the overall temperature profile in a complex manner. This does occur both in semiconducting nanoantennas (e.g. Si, \textbf{Figure \ref{fig:abstract}c}) and in metallic nanostructures (e.g. Au, \textbf{Figure \ref{fig:abstract}b}). However, thermo-optical effects have been largely ignored in thermoplasmonic applications \cite{alabastri_high_2015} and, more broadly, they have never been considered in the modelling of multi-dimensional arrays. Hence, there remains a need for fast approaches to the calculation of photothermal and thermo-optical effects in multi-dimensional arrays of dielectric and plasmonic nanoantennas. 

In this work, we report a fast numerical method to compute the temperature increase of multi-dimensional networks of plasmonic or dielectric nanoparticles embedded in a uniform solid medium, illuminated by a monochromatic light beam \textbf{(Figure \ref{fig:abstract}a}). In particular, our approach leverages a combination of Mie theory and heat-diffusion equations to determine the temperature increase at the location of each nanoparticle (local maxima) as well as in-between nanoparticles (local minima). Importantly, we introduced an energy-conservation based approach, equivalent to the Beer-Lambert law in bulk uniform media, that accounts for the intensity attenuation along the propagation direction due to discrete nanoabsorbers. Thus, going beyond existing solutions \cite{baffou_photoinduced_2013}\cite{baffou_thermoplasmonics_2017},\cite{waiun_parametric_2020}, the proposed method rapidly calculates local extrema of temperature across large multi-dimensional array ($10^5$ particles in less than 1 hour). The obtained values can be also interpolated on specific planes of interest to visualize the photothermal response of the system. In particular, we can trace the transition from a self- to a collective-heating dominated regime for a 3D array of gold nanospheres together with the role of physical parameters such as pitch, illumination intensity and thermal conductivity. Uniquely, by considering a temperature-dependent absorption cross-section of each nanoantenna, our method straightforwardly accounts for thermo-optical effects in these multi-dimensional arrays, without any sensible increase in the required computational time and capacity. Specifically, we show that for both Gold and Silicon nanospheres, under typical laser-level illumination (0.5 $\frac{GW}{m^2}$ ), neglecting thermo-optical effects can lead to an overestimation of the local temperature by several tens of degrees. Interestingly, we also observe that, due to the growing importance of collective heating effects, the impact of thermo-optical changes on absorption is amplified for higher dimensionality arrays. Overall, the proposed method enables, for the first time, a rapid analysis and optimization of both photo-thermal and thermo-optical effects in large multi-dimensional arrays of plasmonic and dielectric nanoantennas alike. While further developments are needed to account for diverse nanoantenna geometries as well as multi-scattering events, we expect that the proposed algorithm will play an important role in the engineering of multi-dimensional arrays with tailored temperature profiles for thermonanophotonic applications\cite{beutel_efficient_2021}\cite{rahimzadegan_comprehensive_2022}.  

\section{Results and Discussion}
\paragraph{}
We address the general case of a homogeneous 3D dispersion of optical nanoantennas embedded in a non-absorbing solid medium and excited by a light beam of radius $R_{beam}$ and uniform intensity $I_0$, propagating along the z-direction  \textbf{(Figure \ref{fig:abstract}a}). In the studied system all nanoantennas are identical, each being characterized by a scattering and an absorption cross-sections, $\sigma_{ext}$ and $\sigma_{abs}$, respectively. Considering the case of spherical nanoparticles, we use Mie theory \cite{baffou_photoinduced_2013}\cite{baffou_thermoplasmonics_2017}\cite{tsoulos_self-induced_2020} to analytically calculate these optical cross-sections (see for example \textbf{(Figure \ref{fig:abstract}b,c}) . Other analytical or numerical approaches (e.g. Gans theory  \cite{eustis_determination_2006}, discrete dipole calculations  \cite{sosa_optical_2003}) can be used to extend the proposed methodology to more complex nanoantenna geometries. The nanospheres are arranged in a network consisting of $N_x, N_y, N_z$ elements along each Cartesian axis. For each dimension, we also define a lattice parameter $\nu_{x,y,z}$ such that $N_{x,y,z}=2\nu_{x,y,z}+1$. As a result, each nanoparticle in the array is associated with a triple of integers $(i, j, k)\in[-\nu_x , \nu_x ]\times[ -\nu_y , \nu_y ]\times[ 1 , 2\nu_z +1 ]$. Similarly to previous works, we finally assume that electromagnetic coupling can be neglected, i.e. nanospheres are separated by center to center distances $P_{x},P_{y}, P_{z}$ that are at least ten times larger than the nanosphere radius $R$ \cite{baffou_photoinduced_2013}\cite{baffou_thermoplasmonics_2017}\cite{kreibig_optical_2005}. 

As previously discussed, the photothermal response of a multi-dimensional array depends on heat generation within each nanosphere, $q_{i,j,k}=\sigma_{abs}I_k$, where $I_k$ is the intensity in the $k^{th}$ plane of the network normal to the propagation beam, as well as heat diffusion across the system. In order to explicitly consider discrete nanoabsorbers, we write an energy balance for the beam before and after each normal plane, obtaining:  
\begin{equation}
    I_k=I_0(1-\frac{\sigma_{abs} N_x N_y}{\pi R_{beam}^2})^{k}
    \label{eq:Idrop_ours}
\end{equation}
This expression, which is analogous to the Beer-Lambert law, is rigorously valid if scattering is negligible compared to absorption. Yet, dominant forward scattering in nanospheres extends its validity to larger nanoresonators (Supplementary Note 1).  

As a result, the temperature increase of each nanoparticle in the network due to its own power absorption (so called self-induced temperature  \cite{baffou_photoinduced_2013}\cite{baffou_thermoplasmonics_2017}), can be expressed as:
\begin{equation}
    \Delta T_{i,j,k}^s=\frac{q_{i,j,k}}{4\pi \kappa R}=\frac{\sigma_{abs}I_0}{4\pi k_{SRM} R}(1-\frac{\sigma_{abs} N_x N_y}{\pi R_{beam}^2})^{k}
    \label{eq:DTs}
\end{equation}
where $k_{SRM}$ is the thermal conductivity of the host medium.

Yet, the total temperature increase of each nanoresonator is equal to:
\begin{equation}
    T_{i,j,k}=T_0+\Delta T_{i,j,k}^s+\Delta T_{i,j,k}^{c}
    \label{eq:T_NP}
\end{equation}
where the temperature increase of each nanosphere due to the thermal energy contribution of its neighbors (so called collective temperature) can now be expressed as:
\begin{multline}
    \Delta T_{i,j,k}^{c}=\frac{I_0}{4\pi k_{SRM}}\sum_{l=0}^{2\nu_z+1}\sum_{m=-\nu_x}^{\nu_x}\sum_{n=-\nu_y}^{\nu_y}\\
    \frac{(1-\frac{\sigma_{abs} N_x N_y}{\pi R_{beam}^2})^{k}}{\sqrt{((m-i)P_x)^2+((n-j)P_y)^2+((l-k)P_z)^2}}\\
    (m,n,l)\neq (i,j,k)
    \label{eq:DTex}
\end{multline}

\begin{figure}[h!]
    \centering
    \includegraphics[width=8cm]{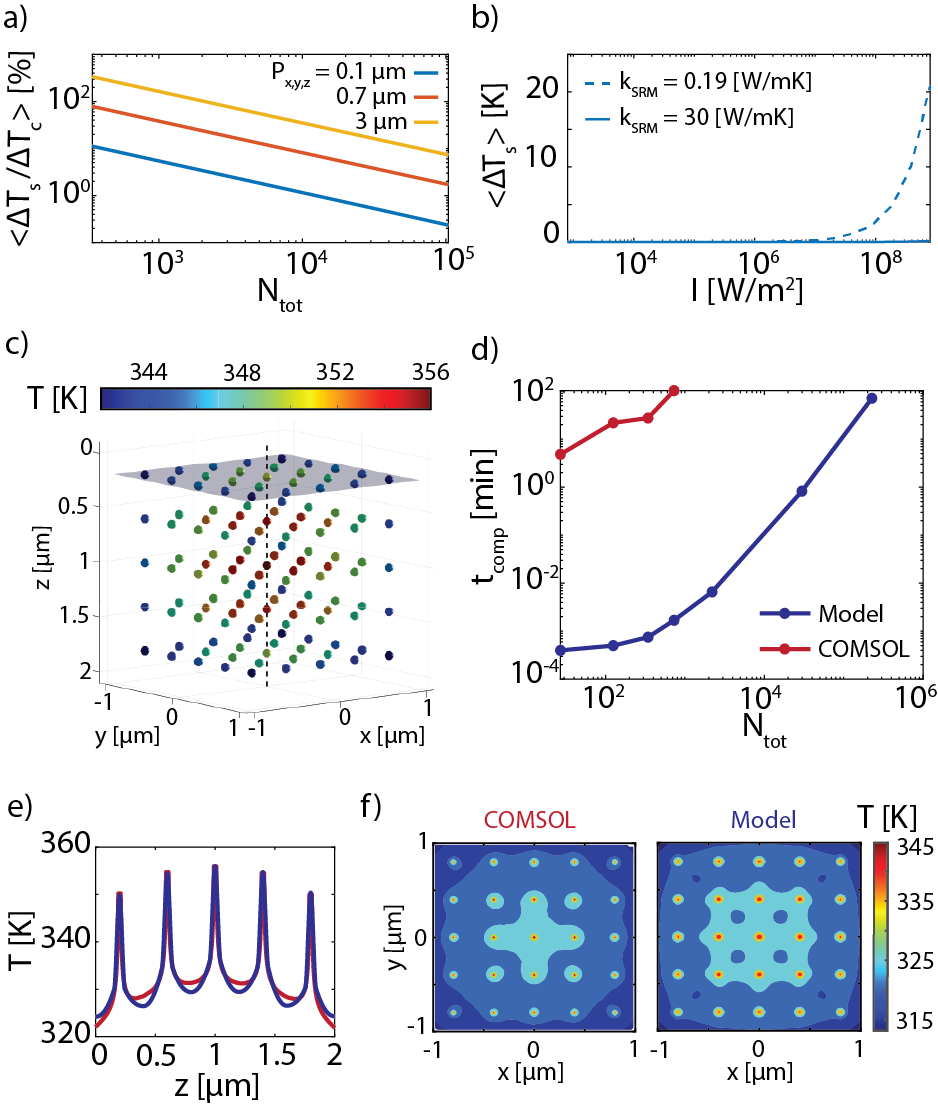}
    \caption{Photothermal Effects in 3D Networks and Model Benchmarking \textbf{a)} Mean ratio of self- and collective heating contributions to temperature (see Eq. \ref{eq:meanDT}) for a 3D isotropic network of increasing size ($N_{tot} = N_x N_y N_z = N_x^3$) as a function of inter-particle distance $P_{x,y,z}$, \textbf{b)} Mean self-heating induced temperature increase of $10^3$ nanoresonators with $P_{x,y,z}=100nm$ embedded in different media for increasing illumination intensity, \textbf{c)} Steady-state temperature of a $[5x5x5]$ network of gold nanospheres ($R = 10 nm$, $P_{x=y=z} = 400 nm$ ) embedded in a medium with $k_{SRM} = 0.19 W/m^2K$ and illuminated with an intensity $I_0 = 1 GW/m^2$ \textbf{d)} Computational time comparison between COMSOL and the proposed model for solving 3D isotropic networks of increasing size, \textbf{e)} Comparison between COMSOL (red curve) and the model interpolation (blue curve) for the temperature profile along z (see dashed line in\textbf{c}) and  \textbf{f)} on the first xy plane (see shaded area in  \textbf{c}).}
    \label{fig:1}
\end{figure}

Therefore, using Eqns (\ref{eq:DTs}) (\ref{eq:T_NP}) (\ref{eq:DTex}) we can compute the temperature at the position $\mathbf{r}$ in the multi-dimensional network by first calculating the temperature increase for each nanosphere, which is inversely proportional to $|\mathbf{r_{i,j,k}}-\mathbf{r}|$, and then summing up all the contributions with the initial temperature. Importantly, we observe that calculating only the local maxima and minima of temperature in the network, corresponding to the location of each nanoabsorber and the mid-way point between absorbers, respectively, is sufficient to retrieve critical information on the heating regime (self versus collective) as well as the range of temperatures to which the material will be subject to. For example, the computed temperature at the location of each nanoabsorber for a $[5x5x5]$ network of gold nanoparticles ($R = 10 nm$, $P_{x=y=z} = 400 nm$, $k_{SRM} = 0.19 W/m^2K$), illuminated with an intensity $I_0 = 1 GW/m^2$, clearly shows the expected temperature localization close to the center of the network (\textbf{Figure \ref{fig:1}c}). As we discuss next, this discrete approach is computationally efficient and is thus well-suited to rapidly investigate the photothermal response of multi-dimensional arrays, including 3D systems. Additionally, we note that ordered networks of nanoparticles can be considered a valid representation of random homogeneous dispersions, broadening the applicability of this approach beyond regular nanostructures. Nonetheless, the code can also be adapted to account for disordered dispersion. (Supplementary Note 2).  

Let's now consider an isotropic 3D network ($N_x = N_y = N_z$) of gold nanoparticles ($R = 10 nm$) embedded in a thermally conductive medium ($k_{SRM} = 30 W/m^2K$) and illuminated with a low intensity of $I_0=1kW/m^2$, comparable to solar irradiation \cite{waiun_parametric_2020}. We also consider three inter-particle distances, $P = 100 nm$, $P = 700 nm$ and $P = 3 \mu m$, corresponding to a dense ($c = 10^{15} cm^{-3}$), sparse ($c = 10^{12} cm^{-3}$) and very sparse ($c = 10^{10} cm^{-3}$) distribution, respectively. In order to follow the evolution from a self- to a collective-heating dominated behavior, we vary the total number of particles in the network ($N_{tot}$) and, for each case, we compute the mean temperature ratio:  
\begin{equation}
    \label{eq:meanDT}
   <\frac{\Delta T^s}{\Delta T^c}> = \frac{\sum_{i=-\nu_x}^{\nu_x}\sum_{j=-\nu_y}^{\nu_y}\sum_{k=0}^{2\nu_z+1} \frac{\Delta T^s_{i,j,k}}{\Delta T^c_{i,j,k}}}{N_{tot}}
\end{equation}
In agreement with previous literature (\cite{richardson_experimental_2009} and \cite{waiun_parametric_2020}), we observe (\textbf{Figure \ref{fig:1}a}) that: (i) collective heating effects increase with the number of nanoparticles; (ii) wider inter-particle separation enhances the self-heating contribution for the same number of irradiated nanoparticles. Specifically, we see that for the dense distribution (blue curve), self-heating effects on average contribute more than 10\% only if the illumination is highly localized (i.e. less than $<10^3$ particles, equivalent to a beam diameter of $1 \mu m$ or smaller). On the other hand, for the sparse distribution (red curve), they still account for more than 10\% of the temperature increase up to $10^4$ nanoparticles (i.e. beam diameter up to $15 \mu m$). For the case of $10^3$ nanoparticles and $P = 100 nm$, we further explore the parameter space and consider the combined effects of illumination intensity and thermal conductivity of the solid medium (\textbf{Figure \ref{fig:1}b}). Interestingly, we observe that, although self-heating effects are small compared to collective ones, under intense illumination ($1 GW/m^2$) and for low thermal conductivities, this contribution results in a mean temperature increase $<\Delta T^s>$ of more than $10 K$. Depending on the application, this contribution might be important, despite being less than 10\% of the total temperature increase. Overall, low intensity and large illumination areas (i.e. uniform solar irradiation) of 3D dispersions result in dominant collective heating effects that can be treated with a uniform heat source, as recently proposed \cite{waiun_parametric_2020}. On the other hand, illumination conditions resembling laser irradiation (sub-mm beam diameter and high intensities, $>500MW/m^2$ ) require a more detailed analysis of the temperature profile, which can be performed efficiently with the proposed method. 

\paragraph{}
In order to benchmark the proposed method, we used COMSOL to perform coupled electromagnetic and thermal calculations of isotropic 3D networks of gold nano-spheres of increasing size and with physical parameters as in \textbf{Figure \ref{fig:abstract}a} (Supplementary Note 3). In addition to significant problems with smooth convergence, the finite element method requires more than one hour of computation for a $[7x7x7]$ network. Additionally, on a desktop computer with a 8-core Intel Xeon W-2145 CPU (3.7$GHz$) and a 128 GB RAM, we could not solve for networks larger than $[9x9x9]$ elements. On the other hand, with the same computational power, the proposed method could compute an array of more than $10^5$ nanoparticles in approximately 70 min (\textbf{Figure \ref{fig:1}d}), greatly expanding our capability to investigate and optimize photothermal effects in large multi-dimensional systems. We also leveraged COMSOL simulations to test the assumptions of our model by changing the nanosphere diameter and separation (Supplementary Note 3).
We confirmed that if the extinction cross-section is dominated by absorption ($R<40 nm$ for Au nanospheres) and electromagnetic coupling can be neglected ($P > 10R$), the model is in excellent agreement with finite element method calculations. 

In order to minimize the computation time, the method focuses only on the local maxima/minima of temperature in the array. Thus, for subsequent visualization of the temperature profile in every point on an arbitrary plane or cut line of the network, we use an interpolation of these values, constrained by a radially decaying temperature profile in the vicinity of each nanoabsorber. \textbf{Figure \ref{fig:1}e} shows the temperature profile along a z-axis cut-line in the center of the 3D array, computed with COMSOL (red curve) and obtained interpolating the discrete temperature values (blue curve). We note that the proposed approach is very accurate in predicting the local maxima while it slightly underestimates the local minima. \textbf{Figure \ref{fig:1}f} additionally show the temperature isocontours on the first plane of the same network, calculated with COMSOL and interpolated from the model, respectively. Despite a reduced accuracy in the regions between each particle, the estimated temperature profile is consistent with the FEM results, with the peak values being in excellent agreement. Thus, our method enables a consistent representation and analysis of photothermal effects in multi-dimensional arrays.  

\begin{figure}
    \centering
    \includegraphics[width=8cm]{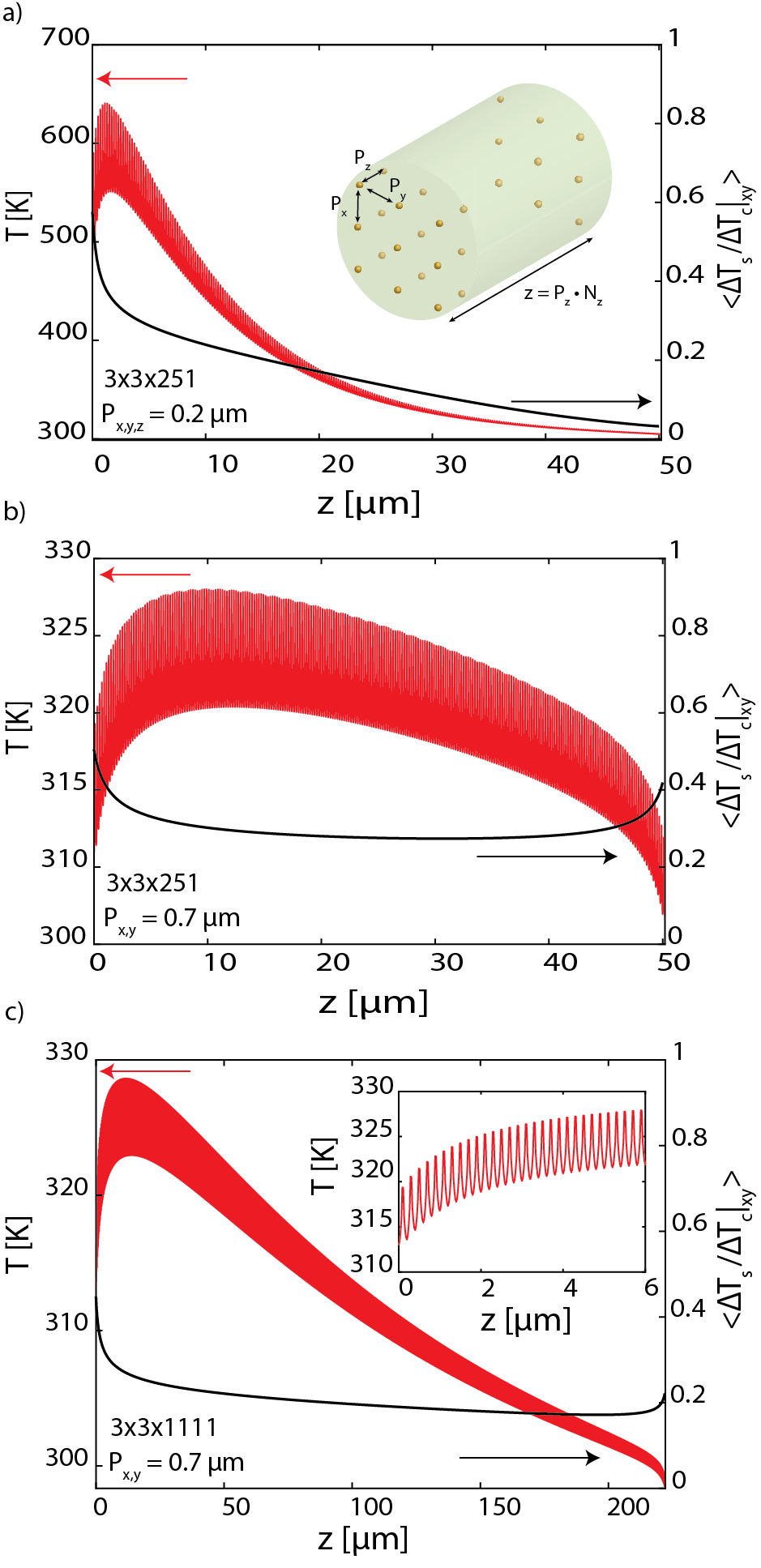}
    \caption{Light Propagation and Temperature Profile \textbf{a)} Temperature profile (red curve, left axis) and relative contribution of self- versus collective heating (black curve, right axis) along z for illumination with $P_{laser}=1\mu W$ of gold nanoparticles ($d=XX$) arranged in a [3x3x251] network with $P_{x,y,z} = 200 nm$, {b)} a [3x3x251] network with $P_{x,y,z} = 700 nm$ (stronger self-heating) and {c)} a [3x3x111] network with  $P_{x,y,z} = 700 nm$. The inset shows a zoom-in of the peaks of temperature due to self-heating.}
    \label{fig:2}
\end{figure}

\paragraph{}
In particular, this approach makes it possible to compute the temperature profile of extended 3D arrays of nanoparticles, including both collective and self heating contributions, without resorting to the uniform heat source approximation. As a case study, we consider gold nanoparticles ($R=10nm$) embedded in a $50 \mu m$ thick slab of low thermal conductivity material, such as PMMA ($k = 0.19 W/m^2$). The interparticle spacing is $P_x = P_y = P_z = 200 nm$, equivalent to a concentration $c=3.6 \cdot 10^{20} cm^{-3}$. A uniform laser beam with total power $1 \mu W$ and diameter $0.567\mu m$ (i.e. $I_0 \approx 4 GW/m^2$ ) is incident on the array. Thus the particles involved in the photothermal effect form an array consisting of [$N_x$, $N_y$, $N_z$] = [3x3x125] elements. \textbf{Figure \ref{fig:2}a} shows the temperature profile of the central line of nanoparticles along the propagation direction, consisting of a continuous background, due to collective heating, and clear temperature peaks, due to self-heating. Interestingly, because of the drop in intensity along the propagation direction, the relative contribution of self-heating, defined as the average of $\Delta T_s/ \Delta T_c$ at each plane xy,  decreases deeper into the array (black line). Indeed, the considered case has a thickness which is larger than the penetration depth, $\delta$, defined as the position where the intensity of the beam is $\frac{I_0}{I_k}=e^{-1}$. For a homogeneous dispersion of nanoantennas in solid media, the effective medium approach \cite{waiun_parametric_2020} leads to:
\begin{equation}
    \label{eq:skd_sivan}
    \delta_{EM}=(c\sigma_{abs})^{-1}
\end{equation}
A similar expression can be obtained from Equation \ref{eq:Idrop_ours} by considering that in our case we modeled the system as a discrete medium
\begin{equation}
    \label{eq:skd_naef}
    \delta_{DM}=-P_zln(1-\frac{\sigma_{abs}N_xN_y}{\pi R_{beam^2}})
\end{equation}
If we define the concentration of the illuminated nanoantennas as follow
\begin{equation}
    \label{eq:c_naef}
    c=\frac{N_{tot}}{V_{Ill}}=\frac{N_xN_yN_z}{A_{beam}N_zP_z}=\frac{N_xN_yN_z}{\pi R_{beam^2}N_zP_z}
\end{equation}
we can reformulate Equation \ref{eq:skd_naef} as 
\begin{equation}
    \label{eq:skd_naef2}
    \delta_{DM}=-P_zln(1-\sigma_{abs}cP_z)
\end{equation}
As shown in \textbf{Figure S3} the two models for the intensity drop (and thus the penetration depth) are equivalent. Yet, the derived expression can account explicitly for inhomogeneous spacing along different spatial directions, in particular along the propagation direction ($P_z$) with respect to the others. 

To observe the effect of the nanoparticle spacing on the temperature profile, we first increase only $P_x$ and $P_y$ from $200 nm$ to $700nm$, while keeping the total incident power constant. As shown in \textbf{Figure \ref{fig:2}b}, by changing the spacing in x and y the temperature along z becomes more uniform. While the magnitude is primarily affected by the reduction in intensity, the temperature profile has changed due to the reduction of the collective heating contribution, that is strongly dependent from the inter-particle distance. This is also confirmed by the nearly constant contribution of self-heating ($<\frac{\Delta T_s}{\Delta T_c}> \approx 30\%$) at all z. The sparser distribution in fact exhibits a penetration depth that is larger than the thickness of the layer. It is important to remark that the calculated temperature profiles intrinsically assume a Dirichlet boundary condition and are thus applicable to situations where convective effects are neglected. This entails that the nanoantennas are not close to an interface but rather surrounded by enough medium to reach the boundary condition by simple conduction. If the thickness of this network is increased along z beyond the penetration depth (3x3x1111 network), a temperature profile similar to the reference case is obtained (\textbf{Figure \ref{fig:2}c}). Yet, the contribution of the self-heating remains higher than in the first case, due to the wider separation between the nanoparticles. Interestingly, we also note that, close to the front and back edges of the network, the self-heating contribution increases due to the reduction in the number of nearest neighbors.  Overall, we observe that the interdependence of the nanoparticle separation (i.e. concentration) on the light attenuation as well as on the relative contribution of self and collective heating can give rise to interesting temperature profiles.

\paragraph{}
Finally, the proposed method allows to seamlessly account for self-induced thermo-optical effects in multi-dimensional arrays. These indeed consist in a change of the complex refractive index of a nanoresonator due to photothermal dissipation. As recently discussed, in Silicon this leads to a spectral shift of the optical resonance (\textbf{Figure \ref{fig:abstract}c}) and ultimately results in off-resonance operation conditions, which considerably impact the final temperature increase of the nanoresonator \cite{tsoulos_self-induced_2020}. Thus, thermo-optical effects must be accounted for by introducing a temperature-dependent absorption cross-section  for each nanoantenna in the network $\sigma_{abs,ijk}(T_{i,j,k})$ \cite{tsoulos_self-induced_2020}. To consider these effects, we modify Equation \ref{eq:Idrop_ours} to consider that, due to thermo-optical effects, the nanoresonators within the network can have different absorption cross sections (for more details Supplementary Note 2)
\begin{equation}
    I_k=I_0\prod_{l=1}^{k}(1-\frac{\sum_{i=-\nu_x}^{\nu_x}\sum_{j=-\nu_y}^{\nu_y}\sigma_{abs,ijk}(T_{i,j,k})}{\pi R_{beam}^2})
    \label{eq:Idrop2}
\end{equation}
where $\sigma_{abs,ijk}$ is the absorption cross section associated with the particle $[i,j,k]$. Then, by modifying  Equations (\ref{eq:DTs}) (\ref{eq:T_NP}) (\ref{eq:DTex}) accordingly and by introducing an iterative loop for the temperature calculation, the proposed method is capable of including this effect. Specifically, based on literature data for the material refractive index, we first compute a temperature-dependent absorption cross-section matrix for the studied nanoantennas (\textbf{Figure \ref{fig:abstract}b,c}). A theoretical or semi-empirical model \cite{alabastri_molding_2013} could also be implemented to broaden the temperature range and the library of materials. Next, the implemented solver computes the temperature of all particles in the network as illustrated previously, using the room temperature absorption cross-section for all nanoantennas. Then, the absorption cross section for each nano-resonator in the network is updated based on the obtained values, $\sigma_{abs,ijk}(T_{i,j,k})$. Finally the new temperatures are computed and the process is repeated until the standard error of the difference in temperatures of the nanoparticles between the iteration $s$ and $s+1$ is below $10^{-4}$ or the absorption cross section for each nano-resonators of the network has been updated to the maximum temperature available in the database storing measurements of the complex refractive index at different temperatures.

To demonstrate the importance of this phenomenon, we have studied the thermo-optical effects in 3D networks of Au and Si nanospheres in solid media ($k_{SRM} = 0.19 W/m^2$). The refractive index $n$ and the extinction coefficient $k$ for the materials at different wavelengths and temperatures have been retrieved from literature \cite{vuye_temperature_1993}\cite{magnozzi_plasmonics_2019}. For this study, we have optimized the radius of the nanoantennas to yield the highest absorption efficiency (i.e. $\frac{\sigma_{abs}}{A_{geom}}$) for both materials (Supplementary Note 4).
We thus consider Au nanospheres with $R=26nm$, whose absorption efficiency as a function of temperature is shown in  \textbf{Figure \ref{fig:abstract}b} It can be observed that in this case temperature-dependent changes in refractive index $n$ and extinction coefficient $k$ reduce the absorption efficiency at the plasmon resonance ($\lambda = 543nm$ at  $25~[^\circ C]$). Hence, as the nanosphere heats up, the absorbed power diminishes and so-does the resulting temperature. Therefore, a temperature overestimation is expected if thermo-optical effects are not included in the calculation, which is typically the case for thermoplasmonic studies. Similar conclusions can be obtained for the case of Si nanospheres. \textbf{Figure \ref{fig:abstract}c}, indeed reports the absorption efficiency spectrum for a $R=33nm$ Si nanosphere. In this case, thermo-optical effects result in both a red-shift and a reduction of the peak in absorption efficiency. Thus, under monochromatic irradiation, initially at the resonance wavelength ($\lambda = 397 nm$),  most Si nanosphere will eventually operate off-resonance, absorbing less power than at room temperature.

\begin{figure}[h!]
    \centering
    \includegraphics[width=8cm]{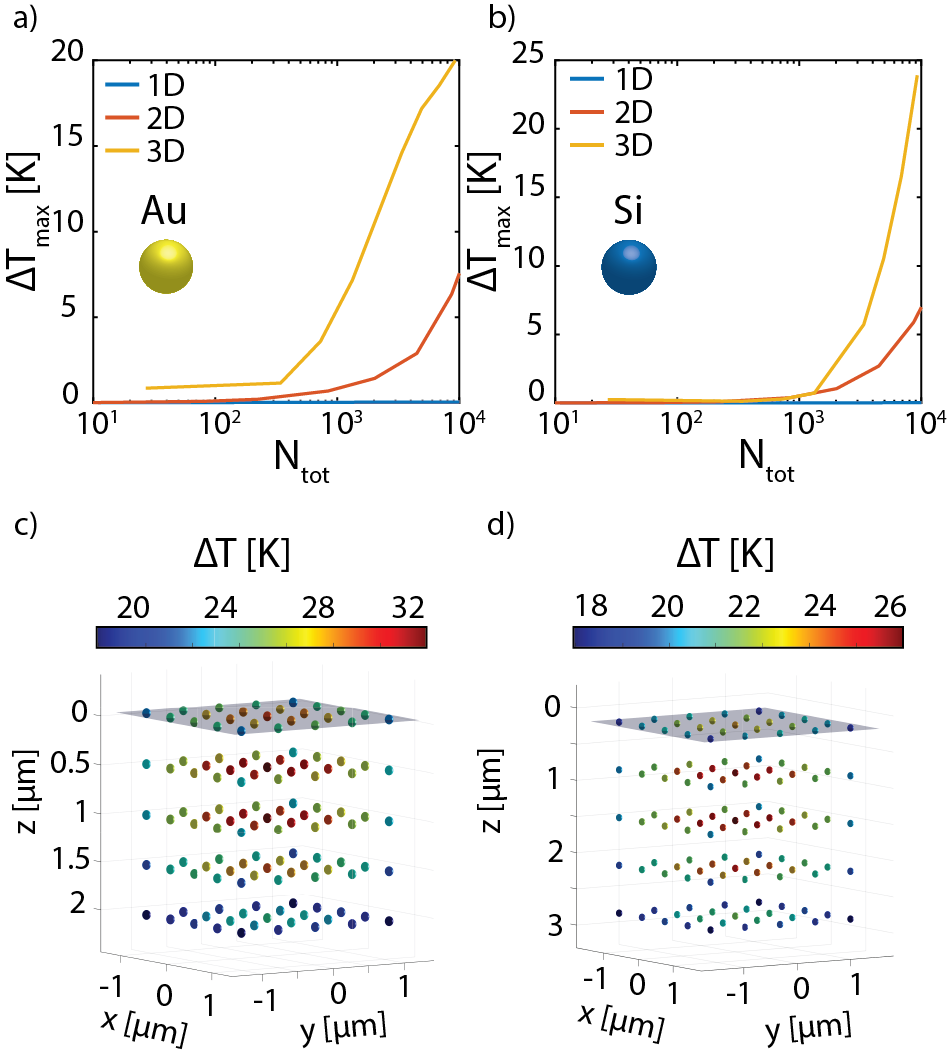}
    \caption{Impact of Thermo-optical Effects on the Photothermal Response of Arrays of Nanoantennas \textbf{a)} Maximum temperature overrestimation due to thermo-optical effects for 1D - [1x1x$N_{tot}$] (blue curve), 2D - [$N_x$ x $N_y$ x 1] (red curve) and 3D - [$N_x$ x $N_y$ x $N_z$] (yellow curve) networks of gold nanosphere with $R=26nm$ and $P_{x,y,z}=520nm$ under illumination at $\lambda=543nm$  and $I_0=50MW/m^2$; \textbf{b)} Same as \textbf{(a)} but for networks of silicon nanopsheres with $R=33nm$ and $P_{x,y,z}=660nm$ illuminated at $\lambda=397nm$  and $I_0=50MW/m^2$; \textbf{c)} Spatially dependent temperature overestimation due to thermo-optical effects in a 5x5x5 network of gold nanospheres with $R=26nm$ and $P_{x,y,z}=520nm$ under illumination at $\lambda=543nm$  and $I_0=500MW/m^2$; {d)} same as {(c)} for silicon nanospheres with $R=33nm$ and $P_{x,y,z}=660nm$ illuminated at $\lambda=397nm$  and $I_0=500MW/m^2$.}
    \label{fig:3}
\end{figure}

While changes in the absorption efficiency directly affects the self-heating contribution to the total temperature increase, the indirect impact of thermo-optical effects on the collective heating contribution is expected to grow as the number of nearest neighbors increases. Thus, the dimensionality of the system must be taken into account. We considered 1-dimensional [$N_x$x1x1], 2-dimensional [$N_x$x$N_y$x1] and 3-dimensional [$N_x$x$N_y$x$N_z$] networks with the same number of total elements ($N_{tot}$) and we calculated the maximum overestimation in temperature if thermo-optical effects are not considered. It is important to note that for network the intensity has been set to $I_0=50MW/m^2$, such that the maximum temperature reached in the system would not exceed the temperature range in the dataset for the temperature dependent complex refractive index of the material (add max value for Au and Si). Figure \textbf{Figure \ref{fig:3}a} and \textbf{\ref{fig:3}b} show that the maximum overestimation in temperature increases both with the total number of nanoparticles and the array dimensionality. This confirms the critical interplay between thermo-optical effects and collective heating, which depends on the number of nanoresonators (i.e. $N_{tot}$) and the closeness of the nearest radiating neighbors. 

To showcase the impact of thermo-optical effects on the spatial temperature profiles we computed a [5x5x5] network of Au nanospheres with $P_{x,y,z}=520nm$ irradiated with  $\lambda = 543 nm$ and $I_0=0.5~[GW/m^2]$.  In this sparse network, self-heating dominates the local temperature increase. \textbf{Figure \ref{fig:3}c} shows the temperature difference obtained for each particle in the network when solving without/with thermo-optical effects. We observe a significant overestimation of the temperature for all nanosphere in the array, ranging from $18K$ closer to the edges and deeper in the network to $30K$ near the center of the network, where collective heating effects are strongest. A similar trend can be observed for a [5x5x5] network of Si nanospheres with $P_{x,y,z}=660nm$  irradiated with  $\lambda = 397 nm$ and $I_0=0.5~[GW/m^2]$ (\textbf{Figure \ref{fig:3}d}). 

Considering that proposed thermo-nanophotonic applications of dielectric nanoresonators often rely on precise phase modulation while thermo-plasmonic device applications rely on target temperature values, a reliable calculation of the local temperature is critical for the overall device functionality. Therefore, if experimentally measured values or models for refractive index $n$ and the extinction coefficient $k$ of the a material are available, the implemented solver is strongly recommended in order to account for thermo-optical effect and to obtain more accurate prediction of the magnitude of the temperature increase within a multi-dimensional array.

\section{Conclusions}
\paragraph{}
In conclusion, in this work we have demonstrated a new solver to model photothermal and thermo-optical effects in multi-dimensional arrays of plasmonic and dielectric nanoantennas embedded in solid media. Extending the capability of previous approaches \cite{baffou_photoinduced_2013}\cite{baffou_thermoplasmonics_2017}\cite{waiun_parametric_2020}, our model calculates the intensity drop along the propagation direction while resolving individual nanoabsorbers. Thus, it can be used to solve for 1D, 2D as well as 3D arrays, seamlessly taking into account the role of nanoparticle spacing (i.e. concentration) on light absorption as well as self- and collective heating effects. Importantly, thanks to its efficient implementation, the model outperforms FEM calculations, breaking the limit of the possible number of simulated nanoparticles while retaining excellent accuracy. Furthermore, by uniquely including a temperature dependent absorption cross-section for each nanoantenna in the array, the proposed methods avoids overestimation of the local temperatures due to thermo-optical effects. Overall, it is a fast and accurate tool that enables the the optimization of photo-thermal effects in multi-dimensional arrays towards a variety of thermo-nanophotonic applications. For example, we believe that an optimized design of 3D arrays could be interesting for realizing thermally-driven self-healing of polymers, glasses and other amorphous materials. In such systems, local temperature gradients could be optimized to minimize the nanoparticle concentration in the medium by balancing the intensity of the beam and the inter-particle distance, while in parallel reaching the local temperature needed for the self-healing of the material. Furthermore, the local temperature profile could be controlled to not surpass a specific upper limit, such as e.g. the melting or glass transition temperature point, in order to avoid altering the desirable properties of the medium material post-healing. This material property-driven optimization could be expanded to and combined with a material-quantity driven optimization towards reduced material aging after healing and global material economy. Furthermore, by leveraging multi-layer structures with different nanoparticle concentrations, it would be possible to take advantage of the temperature profile tunability based on the independent control of the spacing along the xy and z directions. This could be exploited for optimizing the temperature profile for light-actuation devices or in photoelectrochemical systems. Finally, by leveraging computationally-efficient methods to compute the extinction of more complex nanoantennas and account for multi-scattering effects \cite{beutel_efficient_2021}\cite{rahimzadegan_comprehensive_2022}, the proposed approach could be expanded to the design of thermally reconfigurable metasurfaces. Thus, the proposed method introduces a rapid analysis method that will support the development of functional thermo-nanophotonic devices. 

\section{Acknowledgements}
A.N. acknowledges support of the SNSF Eccellenza Grant PCEGP2 194181. T.V.T. acknowledges the support of the SNSF Spark Grant  CRSK-2 190809.

\section{Data Availability}
All software and data are available directly from the authors upon request.

\bibliography{Bibliography}

\end{document}